\documentclass[a4paper,12pt]{article}

\usepackage{amsmath,amssymb,graphics}
\numberwithin{equation}{section}

\newcommand{\ii}{\mathrm{i}}

\newcommand{\dd}{\mathrm{d}}
\newcommand{\pd}{\partial}

\newcommand{\e}{\mathrm{e}}

\newcommand{\tr}{\mathop{\mathrm{tr}}\nolimits}

\newcommand{\I}{\mathbb{I}}

\newcommand{\ft}[2]{{\textstyle\frac{#1}{#2}}}

\begin{document}

\title{Matrix at strong coupling}%
\author{Corneliu Sochichiu\thanks{On leave from~: Institutul de
    Fizic\u a Aplicat\u a A\c S, str. Academiei, nr. 5,
    Chi\c{s}in\u{a}u, MD2028 Moldova; e-mail:
\texttt{sochichi@mppmu.mpg.de}}\\
{\it Max-Planck-Institut f\"ur Physik}\\
{\it (Werner-Heisenberg-Institut)}\\
{\it F\" ohringer Ring 6, D-80805 M\" unchen}
}%

%
%\subjclass{}%
%\keywords{}%

%\date{}%
%\dedicatory{}%
%\commby{}%
% ----------------------------------------------------------------
\maketitle
\begin{abstract}
 We describe the strong coupling limit ($g\to\infty$) for the
 Yang--Mills type matrix models. In this limit the dynamics of the
 model is reduced to one of the diagonal components which is
 characterized by a linearly confining potential. We also shortly
 discuss  the case of the pure Yang--Mills model in more than one
 dimension.
\end{abstract}
\maketitle
% ----------------------------------------------------------------
\section{Introduction}

The development of string and gauge theories is characterized by
their strong inter-relations. The most intriguing result of this
interaction is, probably, the AdS/CFT conjecture
\cite{Maldacena:1997re,Gubser:1998bc} (see \cite{Aharony:1999ti} for
a classical review of the subject). This conjecture relates the
string theory in the Anti-de Sitter background on the one side with
the conformal theory on the Minkowski space-time on the other. The
Minkowski space-time of the conformal theory in this case is related
to the (conformal) boundary of the Anti-de Sitter space.

This conjecture relates a weak coupled model to a
strong coupled one and vice-versa, which is a true Ising-type
duality. Once proved, it would have an 
immense predictive force, e.g. for describing the strong coupled
dynamics of both strings and gauge fields. On the other hand, it is
clear, that for a direct proof, one needs to know the strong
coupled behavior of at least one of these models (in addition to
the weak coupled one for the both). A considerable progress was
achieved in recent years on the way of indirect proofs of the
correspondence (for a recent review see e.g. \cite{Plefka:2005bk}).

On the other hand, in spite of difficulties in the description, it
seems, that the strong coupled regime of the gauge models is the
most natural regime realized in the Nature at the most common (i.e.
low) energies. Perhaps, the most success in the description of the
strong coupled gauge theories was achieved in the framework of the
lattice formulation.\footnote{A good reference for the lattice
approach to gauge theories is given by \cite{Creutz:book}.} An
important problem of this approach, however, is that the continuum
limit of strong coupled
  systems is problematic and it is difficult to separate the
  real physical
effects of the strong coupling from the artifacts of the lattice
description. Therefore, it would be important to have a strong
coupling approach not related to the lattice discretization. In the present
work we attempt to move into this direction.

Although at the end we also consider the Yang--Mills model, the main
subject of this paper is the BFSS type matrix model alias Yang--Mills
mechanics. Yang--Mills type matrix models appear in both contexts of
string and gauge theory. Thus, BFSS \cite{Banks:1996vh} and IKKT
\cite{Ishibashi:1996xs}  matrix models were proposed to describe,
respectively the ``zero''- and ``minus-one''-brane configurations in
the nonperturbative string approach (M-theory). They can be obtained as
dimensional reductions to, respectively, 1+0 and 0 dimensions of the
ten dimensional super Yang--Mills model. (See e.g.
\cite{Sochichiu:2005ex} for a review.)

The plan of the paper is as follows. In the next section we shortly
introduce the matrix model. Then we consider the $g\to\infty$ limit
of the matrix model. First, as a warmup we consider what we call a
strong limit. In this limit we do not consider the contribution from the
high frequency modes. It leads to a model for the diagonal
components where all fields are statistically confined (condensed)
to a single value. Next, we consider a more refined weak limit where
we take into consideration the above higher modes. This leads to a
dynamically nontrivial model for the diagonal components which are
interacting by linear attracting potential. In addition this model
appears to be semi-classical as $g$ goes to infinity. At the end of
this section we discuss a possibility for a systematic expansion at
large coupling.

At the end we discuss the possibility to extend the analysis to the
Yang--Mills model.

% ----------------------------------------------------------------
\section{Matrix model}

Consider the matrix model (Yang--Mills mechanics) which is
described  by the following classical action:
\begin{equation}\label{action}
  S=\int \dd t \tr\left\{\ft12
  (\nabla_0X_a)^2+\ft{g^2}{4}[X_a,X_b]^2\right\},
\end{equation}
where $X_a$ are $D$, $a=1,\dots D$ time dependent Hermitian $N\times
N$ matrices while $g$ is the gauge coupling. The covariant time
derivative is defined
by the use of the (non-dynamical) temporal gauge field $A\equiv A_0$,
\begin{equation}
  \nabla_0X_a=\dot{X}_a+[A,X_a].
\end{equation}
The role of the gauge field is to impose the Gauss law constraint
$[X_a,\nabla_0 X_a]=0$ which provides the gauge invariance of the action with
respect to the time-dependent U($N$) gauge transformations,
\begin{equation}\label{gauge}
  X_a\mapsto U^{-1}(t)X_a U(t), \qquad A\mapsto
  U^{-1}(t)AU(t)+U^{-1}\dot{U}
\end{equation}
where $U(t)\in $U$(N)$. Other features of the model include:
\begin{itemize}
  \item Invariance with the respect to shifts by a constant scalar
  matrix
    \begin{equation}\label{shift}
      X_a\mapsto X_a+c_a\cdot\I.
    \end{equation}
Restricting the gauge group to SU$(N)$ removes this degree of freedom
  \item Invariance with respect to the (target space) rotations,
    \begin{equation}\label{lorenz}
      X_a\mapsto \Lambda_a{}^b X_b,
    \end{equation}
$\Lambda\in $SO$(D)$
  \item In the case of $D=10$ eq. \eqref{action}
represents the bosonic part of the supersymmetric BFSS matrix model
\cite{Banks:1996vh},
\begin{equation}
  \int \dd t \tr\left\{\ft12
  (\nabla_0X_a)^2+\ft{g^2}{4}[X_a,X_b]^2+\psi\nabla_0\psi+
  \psi\Gamma^a[X_a,\psi]\right\},
\end{equation}
where $\psi$ is the fermionic $N\times N$ matrix with 10 dimensional
Majorana--Weyl fermionic indices.
\end{itemize}

For the matrix model under consideration one can formulate a
perturbative expansion in terms of the powers of the gauge coupling
$g$ similar to the perturbative expansion of the Yang--Mills
theory. In what follows we will not discuss this type of perturbative
expansion but refer the reader to the appropriate
Yang-Mills perturbation theory literature instead.

% -----------------------------------------------------------------
\section{Spontaneous symmetry breaking at strong coupling}

It is expected that the
strong coupling limit $g\to\infty$ implies the commutativity of the
matrices $X_a$,
\begin{equation}
  [X_a,X_b]=0.
\end{equation}
Indeed, as $g$ goes to infinity the path integral contribution of
configurations with non-zero commutator are exponentially
suppressed.

Since this is the case, one can diagonalize simultaneously all the
matrices $X_a$, whose eigenvalues would then correspond to the
coordinates of some branes. In this case one can say that in the strong
coupling limit the branes can be localized. (Beyond this limit they
are fuzzed by the strings by which the branes interact.)

Let us consider the above $g\to\infty$ limit in more details. For this
let us
split the matrix degrees of freedom $X_a^{mn}$ into the diagonal part:
\begin{equation}
  x_a=\mathop{\mathrm{diag}}{X_a},
\end{equation}
and the remaining off-diagonal one:
\begin{equation}
  z_a=X_a-x_a.
\end{equation}

This splitting seems somehow abusive, since it does not respect the
gauge invariance \eqref{gauge}. In fact, it corresponds to a
particular choice of commutative background among gauge equivalent
ones. This choice  breaks \emph{spontaneously} the U$(N)$ symmetry
down to U(1)$^N$. At the same time $z_a$ can be treated as a
perturbation above this background.

The spontaneous breaking of
the symmetry is always associated with the zero modes corresponding to
different gauge equivalent choices of the background\footnote{In the
  present case this is the symmetry: $x_a\to U^{-1}x_aU$.}.
An appropiate choice of SU$(N)$ gauge apparently solves this problem
since it restricts the allowed perturbations of the vacuum to the
transversal direction. The unbroken gauge symmetry as well as the
possibility to fix the gauge depends strongly on whether the
diagonal background is degenerate or no. Although the exceptional
configurations with the degenerate background may in principle
contribute (and even dominate) in spite of zero measure, we so
far neglect this issue and consider in rest of this paper the
general position point: where all diagonal eigenvalues $x_n$ are
different (as $D$-dimensional vectors).

% -----------------------------------------------------------------
\section{$g\to\infty$: the strong limit}

On can define different strong coupling limits depending on the
relation of the coupling with other parameters (like $N$ or the
cut-off). In this section we consider the \emph{strong limit}: This
limit assumes that the model is UV-regularized and the limit
$g\to\infty$ is taken prior to removing the regularization.
Technically, this means that one can drop in this limit the time
derivatives if they come with a factor vanishing in the limit
$g\to\infty$. In contrast to this, the \emph{weak limit} which is
taken after the removal of (or eventually not imposing at all) the
regularization is discussed in the next section.

In the non-degenerate case the whole U$(N)$ gauge group is broken by
the diagonal component of the background down to U$(1)^N$. The
infinitesimal gauge transformation of the background is given by
$\delta x_a=0$ and $\delta z_a=[x_a,u]+[z_a,u]$. This is very
similar to the ordinary gauge transformation in the nonabelian
Yang--Mills theory if the role of the partial derivative operator
$\pd_a$ is attributed to the commutator $[x_a,\cdot]$. In the
complete analogy with this one can fix the gauge by imposing the
Lorenz gauge condition\footnote{Admissible gauge fixing and
  corresponding Faddeev--Popov determinants are discussed in the
  classical book on gauge theories \cite{Slavnov:book}.}:
\begin{equation}\label{str-gauge}
  F_{\rm g.f.}\equiv [x_a,z_a]=0.
\end{equation}
The Faddeev--Popov determinant corresponding to the gauge fixing
condition \eqref{str-gauge} is given by
\begin{equation}
  \Delta^{(\infty)}_2(x)=
  \prod_{{\rm time}}\left[{\prod_{mn}}'(x_a^m-x_a^n)^2\right]^{\ft 12},
\end{equation}
where the prime denote that the product extends over the distinct
indices $m$ and $n$ only. Formally, the determinant is different
from zero (which is important for the implementation of the gauge
condition) when all $x$-eigenvalues are given by distinct points
$x_a^n$, $n=1,\dots N$.

All above can be appropriately formalized in the quantum theory by
adding the gauge fixing term and the Faddeeev--Popov determinant in the
(Euclidean) partition function which takes the form
\begin{equation}
  Z=\int [\dd x][\dd z][\dd A]\, \Delta^{(\infty)}_2 (x)\exp\{-(S+\ft
  {g^2}2
  \tr[x_a,z_a]^2)\},
\end{equation}
where we used so called ``alpha-gauge'' (with $\alpha =g^2$)
implementation of the gauge fixing rather than the ``delta-function
implementation''. Note, that since the introduction of the gauge
fixing condition \eqref{str-gauge} one cannot anymore impose any
further restriction\footnote{Except for the
  vanishing of the diagonal part of $A$, $A_{nn}=0$.}
on the gauge field $A$ which should remain in the action.

Now we are ready to take the limit $g\to\infty$ and separate the
leading contribution in this limit. There are several ways to do this,
which, naturally, lead to the same result. Let us consider the
following one. Let us substitute the variables $z_a$ by the rescaled ones
as follows
\begin{equation}
  z_a\to g z_a.
\end{equation}
Then, the matrix action \eqref{action} takes the following form:
\begin{multline}\label{x-z-action}
  S=
  -\int\dd t\,
  \left(\ft12 \dot{x_a}^2+\ft{1}{2g^2}
  \dot{z_a}^2+\ft 1g[A,z_a]\dot{x}_a+\ft 12 [A,(x_a+\ft 1g
  z_a)]^2+\right.\\
  \left.\ft1g[A,(x_a+\ft 1gz_a)]\dot{z}_a+\ft
  14([x_a,z_b]-[x_b,z_a]+\ft{1}{g^2}[z_a,z_b])^2+\ft12 [x_a,z_a]^2]
  \right).
\end{multline}
As we are taking the strong $g\to\infty$ limit, we should discard all
terms formally vanishing in this limit. Thus, the leading part of the
action becomes
\begin{equation}
  S_{g\to\infty}= -\int\dd t\,
  \left(\ft12 \dot{x_a}^2+\ft 12 [A,x_a]^2+\ft
  12[x_a,z_b]^2\right).
\end{equation}

The action is quadratic in the gauge field $A$ as well as in the
off-diagonal field $z_a$. Integrating in both $A$ and $z_a$, one gets
the factor coinciding with the Faddeev--Popov determinant at the power
$-(D+1)/2$. The partition function then reads,
\begin{equation}\label{gppp}
  Z=\int\left[\dd x\,\Delta^{(\infty)}_2(x)^{-(D-1)
  /2}\right]\exp\left\{\int\dd
  t\,\ft12\dot{x}^2\right\},
\end{equation}
which appart from the determinant factor in the measure corresponds to
a free particle partition function.

The modification of the measure in \eqref{gppp} signals the confining
of the eigenvalues $x_n$ to a common value which itself is a subject
to free motion. Indeed, in the case of only two eigenvalues the path
integral \eqref{gppp} reduces to (see the Appendix \ref{zwei}),
\begin{equation}\label{z-4-2-a}
  Z=\int[\dd^D Y]\,[\dd^D y\, y^{-2(D-1)}]\, \e^{\ii\int \dd t \left(\ft
  12\dot{Y}^2+\ft12 \dot{y}^2\right)},
\end{equation}
where $y$ is the distance between branes while $Y$ is the free
moving ``center of mass''. Consider the $y$-measure locally at the
instant $t$: $\dd^D y (y^{2})^{-D+1}(t)=\dd \Omega_D\dd r r^{-D+1}$.
Integration with such a measure is divergent at
$r\equiv\sqrt{y^2}=0$ unless the integrand vanishes quickly enough
as $y$ approaches the origin, which is not the case for slow $y$
modes. Statistically this means that configurations with small $y^2$
produces a contribution to the partition function which is
infinitely larger than the contribution of all the configurations
with larger values of $y^2$. Therefore, under the normalization the
configurations with \emph{nonzero} $y^2$ will get \emph{zero}
expectation values. One can see also that the conclusion is very
sensitive to the power of $\Delta_2^{(\infty)}$. Thus, if the power
were e.g. $-(D-1)/4$ no such statistical confinement would occur.

It may appear however that this simple estimation of $g\to\infty$ is
too rough and one must weaken the limit allowing the contribution of
higher frequency modes. We come to this in the next section.

% -----------------------------------------------------------------
\section{$g\to\infty$: the weak limit}

Consider the stationary points of the action \eqref{action} i.e. the
solutions to the equations of motion.
There is a class of static solutions to the equations of motion  given
by constant commuting matrices
$x_a$. We can assume that these matrices depend adiabatically on time.
One can consider perturbations about this background.
The perturbation is given by the off-diagonal part $z_a$ as well as by
the fast diagonal modes. The diagonal modes do not contribute at the
one-loop level since there are no nonlinear terms in the action
corresponding to diagonal-diagonal interaction. As a sequence, we can
neglect the fast diagonal fluctuations and consider only the adiabatic
modes.

Therefore, consider the contribution of the off-diagonal modes as well
as of the auxiliary (gauge and ghost) fields and evaluate their
contribution in the one-loop approximation in $1/g$
expansion. Throughout this section
we use the Euclideanized version of the theory.

To proceed with the evaluation let us fix the gauge by adding the
following gauge fixing term to the Lagrangian:
\begin{equation}\label{gf}
  L_{g.f.}=-\tr \left(\ft{1}{2g^2}\dot{A}^2+\ft12[x_a,z_a]^2\right).
\end{equation}
The variation of the gauge fixing condition gives the
Faddeev--Popov operator,
\begin{equation}\label{fp}
  M_{FP}u=\pd_0\nabla_0u+[x_a,[(x_a+z_a),u]],
\end{equation}
whose determinant $\Delta_{FP}=\det M_{FP}$ is the
\emph{Faddeev--Popov determinant} which we have to use together with
the condition \eqref{gf}.
In the one loop approximation no contribution
will come from $A$- and $z$-dependent terms in the Faddeev--Popov
operator. Therefore, in what follows we will discard
these terms. As a result, the Faddeev--Popov determinant restricted to
one loop relevant terms takes the
following form
\begin{equation}\label{delta2}
  \Delta_{FP}|_{\rm
  (1~loop)}={\prod_{m,n}}'\det\left[-\ft{1}{g^2}\pd_t^2+r_{mn}^2\right]
  \equiv \Delta_2(x),
\end{equation}
where $r_{mn}^2=(x^a_m-x^a_n)^2$ is the square distance between $n$-th
and $m$-th branes and the prime denotes that the product is taken for
distinct $m$ and $n$.

Let us turn to the action \eqref{action}. The matrix model action can
be rewritten in the form as follows,
\begin{multline}\label{1l-act}
  S=-\int\dd t\left(
  \ft12 (\dot{x}^a_n)^2+\ft1{2g^2}|\dot{z}^a_{mn}|^2
  +\ft1{2g^2}|\dot{A}_{mn}|^2+
  \ft1{g^2}\dot{\bar{c}}_{mn}\dot{c}_{mn}\right. \\
  \left.+\ft12 r_{mn}^2 (|z^a_{mn}|^2+|A_{mn}|^2+\bar{c}_{mn}c_{mn})
  +\dots\right)
\end{multline}
where the dots stand for the terms not contributing at the one loop
level
(e.g. terms which are higher than the second order in $A$ and $z$).

After the integration over the gauge field $A$, the off-diagonal
component $z$ and the
ghosts $c$ and $\bar{c}$ the partition
function takes the form,
\begin{equation}\label{z-det}
  Z=\int[\dd x]\Delta_2^{-\ft{D-1}{2}}(x)\,\e^{\int\dd t
  \,\ft12\dot{x}^2 }
\end{equation}

As it can be seen, the problem is reduced to the computation of the
determinant $\Delta_2$, of an elliptic differential operator.
Let us use the $\zeta$-function
approach to do such a computation\footnote{A similar computation for
  constant diagonal modes was done in \cite{Fatollahi:1999na} and its
  phenomenological implications were explored in \cite{Fatollahi:2001dz}. I
thank Amir H. Fathollahi for pointing my attention to these papers.}
 (see e.g. \cite{Schwarz:book}). According to
this approach, the logarithm of the determinant of an elliptic operator
$D$ is given by the (minus) derivative of the $\zeta$-function,
\begin{equation}
  \ln \det D=-\zeta'_D(0),
\end{equation}
where the function $\zeta_D(s)$ is defined as  the analytic
continuation of the series,
\begin{equation}
  \zeta_D(s)=\sum_\lambda \frac{1}{\lambda^s}
   =\frac{1}{\Gamma(s)}\int_{0}^{\infty}
   \dd \rho \rho^{s-1}\tr\e^{-\rho D}.
\end{equation}

The trace  $\tr\e^{-\rho D}$ can be written as
\begin{equation}\label{tr-ker}
  \tr\e^{-\rho D}=\int\dd t\, K_D(t,t;\rho),
\end{equation}
where $K_D(t',t'';\rho)$ is the Heat Kernel for the operator $D$. It
is the solution to the Heat Equation
\begin{equation}\label{he}
  \pd_\rho K(t,t_0;\rho)=-DK(t,t_0;\rho),
\end{equation}
with the initial conditions given by
\begin{equation}
  K(t,t_0;0)=\delta(t-t_0).
\end{equation}

In the case at hand $D=\left(-\ft1{g^2}\pd_t^2+r_{mn}^2\right)$ and
the solution for the Heat Kernel is given by
\begin{equation}\label{hk}
  K(t',t'';\rho)=
  \frac{g}{\sqrt{4\pi\rho}}\exp\left(-\frac{g^2(t''-t')^2}{4\rho}
  -r_{mn}^2\rho\right) .
\end{equation}

Since the time integral in the r.h.s of the equation \eqref{tr-ker}
diverges for $t\in (-\infty,+\infty )$ it is useful to put the system
in the time box interval $\tau$. Beyond its regularization function
the $\tau$ plays another important role, namely, that of being also
the \emph{adiabaticity} box. Roughly speaking, the $\tau$-interval is
the ``$\dd t$'' for the adiabatic time ``$t$''.

The $\zeta$-function for the time interval $\tau$ is then given by
\begin{multline}\label{zeta-s}
  \zeta_D(s)=\frac{g\tau}{\sqrt{4\pi}\Gamma(s)}{\sum_{mn}}'
  \int_{0}^{+\infty}
  \dd\rho\rho^{s-3/2}\,\e^{-r_{mn}^2\rho}\\
   =
  \frac{g\Gamma(s-1/2)}{\sqrt{4\pi}
  \Gamma(s)}\tau{\sum_{mn}}'(r_{mn}^2)^{1/2-s}.
\end{multline}

Computing the derivative of \eqref{zeta-s} and taking the limit $s\to
0$ we obtain:
\begin{equation}
  -\zeta_D'(0)=g\tau{\sum_{mn}}'\sqrt{r_{mn}^2}.
\end{equation}

Summing over the all adiabatic boxes we get:
\begin{equation}\label{potential}
  \Delta_2^{-\ft{D-1}{2}}(x)=\e^{-\frac{g(D-1)}{2}\int\dd t
  \sum_{mn}\sqrt{r_{mn}^2}},
\end{equation}
where we can even drop the prime from the sum.

Therefore, the low energy effective action for $x_n^a$ takes the form,
\begin{equation}\label{final}
  S_{g\to\infty}=\int\dd t\,\left(-\ft12 \dot{x}_n^2
  -\ft12 g(D-1)\sum_{mn}\sqrt{(x_m-x_n)^2}\right).
\end{equation}

As one can see, the action \eqref{final} corresponds to a system with
strong linear confinement of the particles. In spite of its terrifying
appearance the limit $g\to\infty$ corresponds to nothing else then the
semi-classical limit. Indeed, passing to a rescaled $x_n^a$,
\begin{equation}
 x^a_n\to x^a_n/g(D-1),
\end{equation}
transforms the partition function \eqref{final} to the following
semi-classical form
\begin{equation}\label{finalest}
  Z=\int[\dd x]\e^{g^2(D-1)^2\int\dd t\,\left(-\ft12\dot{x}_n^2
  -\ft12\sum_{mn}\sqrt{(x_m-x_n)^2}\right)},
\end{equation}
where $g^2$ plays the role of inverse Planck constant
$\hbar^{-1}$. In fact, the above rescaling introduces a
renormalization of the brane
coordinate. Its meaning is that the nontrivial dynamics corresponds to
large (in the old scale) brane separations. Therefore, the natural scale of
the brane dynamics is given in terms of the attraction force (tension)
acting on the branes.
\subsubsection*{A remark on the systematic expansion}
A trick can be used to modify the value of the coupling constant
(and even to invert it). 

We can consider the model at the finite
temperature $T=1/\beta$. The finite temperature implies that the
action in the path integral is computed for the Euclidean time
interval $0\leq t<\beta$ with periodical boundary condition for the
fields. A simple dimensional analysis that the following rescaling
\begin{equation}\label{scal}
  \beta\to\beta/\lambda^2, \qquad g^2\to g^2\lambda^6,
\end{equation}
changes the partition function by a constant multiplicative factor
only, which can be absorbed in the measure. Indeed, making the
substitution $X\to\lambda X$ one gets \eqref{scal}. Now taking
$\lambda$ arbitrarily small one can make $g$ small as well, e.g.
equal to $g^{-1}$. At the same time $\beta$ goes to infinity i.e. the
theory rolls down to zero temperature.

Unfortunately, because of different scaling
properties, this trick can not be used in the case of Yang--Mills
theory in more than two dimensions.
% ---------------------------------------------------------
\section{The Yang--Mills model}

It is tempting to apply the above analysis to the SU($N$)
Yang--Mills model. Let us enumerate the modifications that occur
when passing to the pure $D$-dimensional Yang--Mills model:
\begin{itemize}
   \item Instead of
the determinant
\eqref{delta2} one should compute the determinant of the
$D$-dimensional differential operator
\begin{equation}
  D=\ft1{g^2}\pd_\mu^2-r_{mn}^2,
\end{equation}
where the diagonal modes are described by the Abelian gauge fields
$a_\mu^n(x)$, $r_{mn}^2=(a_\mu^m-a_\mu^n)^2$. Also since the gauge
group is SU$(N)$ the center of mass is fixed:
\begin{equation}\label{cm}
\sum_n a^n_\mu=0.
\end{equation}
  \item Heat Kernel:
    \begin{equation}
      K(x',x'';\rho)=\frac{g^D}{(4\pi\rho)^{D/2}}
      \exp\left(-\frac{g^2(x''-x')^2}{4\rho}-r_{mn}^2\rho\right).
    \end{equation}
   \item The one loop contribution is given by:\footnote{The factor
   $(D-2)$ instead of
   $(D-1)$ as in the previous sections appears
   because the gauge field $A_0$ is now counted as one of the fields.}
     \begin{equation}\label{s-eff}
       L_{eff}=\ft14
       (F_{\mu\nu}^n)^2-g^D(D-2)\sum_{mn}V_{mn}(a),
     \end{equation}
   where $F_{\mu\nu}^n=\pd_\mu a^n_\nu-\pd_\nu a^n_\mu$ and
   \begin{equation}
     V_{mn}=\frac{(r_{mn}^2)^{D/2}}{(4\pi)^{D/2}}\times
     \begin{cases}
       \Gamma (-D/2),& D\text{-odd}\\
       \frac{(-1)^{D/2}}{(D/2)!}\left(\log r_{mn}^2-\mathop{h}(D/2)\right),
       & D\text{-even}
     \end{cases}
   \end{equation}
   where $h(k)$ is the $k$-th harmonic number: $h(k)=\sum_{l=1}^k
   1/l$. (Note also that the $\Gamma$-function is regular at negative
   half-integer points.)
\end{itemize}

As it could be seen, for $D>2$ ($D=2$ is dynamically trivial) one can
rescale the fields
\begin{equation}
  a_\mu^n\to g^{\ft{D}{D-2}}a_\mu^n,
\end{equation}
and get a common factor $g^{\ft{2D}{2-D}}$ in front of the effective
action. For $D>2$ this factor vanishes in the limit $g\to\infty$,
which means that in this case the quantum fluctuations are strong.
As one can note, the the qulitatative behaviour of the effective
models depends on dimension. Thus, in dimensions $D=4k$ and $4k+1$ for
non-negative integer $k$, the strong attractive force binds al $a^n$
together, while for $D=4k+2$ and $4k+3$ the repulsive force keeps them
appart at infinity. The common feature is that in
this situation we are not able to catch any nontrivial dynamics beyond 
the fact that all diagonal values are confined to zero or infinity.

A much more serious problem is that for $D>2$ the higher loop
contribution is not suppressed at large $g$ unless an UV-cutoff
($\Lambda<\infty$) is used. A nontrivial contribution can be then
catched taking the double scaling limit with $g\to\infty$ and
$\Lambda\to\infty$. A more detailed analysis would give the answer
whether this is possible.

% ---------------------------------------------------------
\section{Discussion}

In this paper we considered the strong coupling limit of the matrix
model. It is shown that the modes which survive in this limit are
described by a system of linearly interacting particles. As coupling
goes to infinity the system becomes semi-classical $g^2$ playing the
role of inverse the inverse Planck constant $\hbar^{-1}$. The scale
at which the dynamics takes the semi-classical form is given then by
the string tension or the coefficient of linear interaction. The
analysis is performed at the one-loop level. It seems rather
possible that a systematic expansion in the inverse powers of the
coupling constant can be constructed in addition to the standard
small coupling constant expansion.

It is also very tempting to apply the $1/g$-expansion to the
Yang--Mills theory. The one-loop technique can be easily extended to
the ordinary Yang--Mills model. In the case of the dimensionality
higher than two the diagonal component is not anymore semi-classical
and most probably does not decouple. The implications of this  are
not yet clear. There are however, resources we did not use which are
given by the large $N$ and UV cut-off scaling. Taking a correlated
limit of large $g$, $N$ and $\Lambda$ one may hope to get a
non-trivial content for the expansion, e.g. by tuning the
background.

Another important issue we left beyond our consideration regards the
exceptional configurations with some $r_{mn}=0$. As the effective
parameter of the expansion is $1/gr_{mn}$ the expansion fails if
some $r_{mn}\lesssim g^{-1}$. Important point is the statistical
weights of such configurations. An estimate can be done by the
computation of the average separation $\bar{r}$. When the average
separation is nonzero $\bar{r}>0$, it is clear, that one can trust the
approach. In
the case of pure Yang--Mills model, however, it seems that it
either vanishes or is infinite according to the dimension.

In the case of branes at close distance the $2\times 2$ matrix block
corresponding to respective eigenvalues is not decoupled and one
should consider the entire matrix dynamics similarly to what is done
in the non-commutative gauge theory
\cite{Sochichiu:2001am,Sochichiu:2001kb,Sochichiu:2001kc}. As it was
found this dynamics is a stochastic one.

% ---------------------------------------------------------
\subsection*{Acknowledgments}
I thank Dieter Maison and Peter Weisz for useful discussions.
This work was done under the Humboldt fellowship program.

\appendix
\section{Example: The tale of two branes}\label{zwei}

Consider the case of two branes. In this case the action can be
written in the following form:
\begin{multline}\label{su2action}
   \mathcal{L}=\ft12\dot{Y}^2+\ft12\dot{y}^2+\dot{z}\cdot
   \dot{\bar{z}}\\
   +\sqrt{2} \left[a (\dot{y}\cdot
   \bar{z}-
   \dot{\bar{z}}\cdot y)-  \bar{a}(\dot{y}\cdot z
   -\dot{z}\cdot y)
   \right]-\left(a_1-a_2\right)\left(\dot{z}\cdot \bar{z}
   -\dot{\bar{z}}\cdot
   z\right)
   \\
   +\sqrt{2}\left(a_1-a_2\right)( a y\cdot
   \bar{z}+ y\cdot z \bar{a})-2 a\bar{a} (z\cdot \bar{z}
   +y^2)+z^2 \bar{a}^2+a^2
   \bar{z}^2\\
   -\ft{1}{2} z\cdot \bar{z}
   \left(a_1-a_2\right)^2\\
   -g^2\left(2 y^2 z\cdot \bar{z}+\left(z\cdot
   \bar{z}\right)^2-z^2 \bar{z}^2\right)
   -2y^2(c\bar{c}+c^*\bar{c}^*),
\end{multline}
where $2\times 2$ matrix $X_a$ is given by the following component
structure,
\begin{equation}
  X_a=
  \begin{pmatrix}
    \frac{1}{\sqrt{2}}(Y_a+y_a)& z_a\\
    \bar{z}_a &\frac{1}{\sqrt{2}}(Y_a-y_a)
  \end{pmatrix},
\end{equation}
while the gauge field matrix $A$ is given by the components:
\begin{equation}
  A=
  \begin{pmatrix}
    a_1& a\\
    \bar{a} & a_2
  \end{pmatrix},
\end{equation}
and two complex conjugate components of the ghost-anti-ghost are used. 
All diagonal components are real while the off diagonal elements are
complex. The dot in \eqref{su2action} indicates the inner product
with respect to the index $a=1,\dots,D$.

The first four lines of \eqref{su2action} are the contribution from
the kinetic term while
the third line comes from the commutator term together with the gauge
fixing term and Faddeev--Popov determinant for the gauge $[x,z]=0$.

Let us make the following substitution:
\begin{equation}
  z\to g z.
\end{equation}
After the rescaling one can split the Lagrangian \eqref{su2action} in
the leading term and perturbation in $1/g$. The leading part of the
Lagrangian looks as follows,
\begin{equation}\label{g-inf}
  \mathcal{L}_{g\to\infty}=\ft12\dot{Y}^2+\ft12\dot{y}^2
   -2 y^2 (a\bar{a}
   + z\cdot \bar{z}+c\bar{c}+c^*\bar{c}^*).
\end{equation}

All fields with the exception of the $y$ and the free $Y$ become
non-dynamical in the limit $g\to\infty$ and the Lagrangian
\eqref{g-inf} is quadratic this fields. Therefore, integration of
$z$, $\bar{z}$, $c$, $\bar{c}$ and $a$ leads\footnote{Remaining
components contribute by only a
  constant factor.} to the partition function of the form
\eqref{gppp},
\begin{equation}\label{z-4-2}
  Z=\int\prod_t \dd^D Y\,d^D y\, [y^2(t)]^{-D+1}\, \e^{\ii\int \dd t
  \left(\ft
  12\dot{Y}^2+\ft12 \dot{y}^2\right)}.
\end{equation}

As in the case of \eqref{gppp} the measure in eq. \eqref{z-4-2} is
singular as $y^2\to 0$.

% -----------------------------------------------------------------

% ----------------------------------------------------------------
% ----------------------------------------------------------------
%\bibliographystyle{h-elsevier}
%\bibliography{lect}
% ----------------------------------------------------------------

\end{document}